\documentstyle[preprint,aps]{revtex}
\tightenlines
\def\be{\begin{equation}}
\def\ee{\end{equation}}
\def\bea{\begin{eqnarray}}
\def\eea{\end{eqnarray}}
\def\ba{\begin{array}}
\def\ea{\end{array}}
\def\a{\alpha}
\def\b{\beta}

\def\d{\delta}

\def\0{$\Gamma_0$}

\def\t{\tau}

\def\s{\sigma}

\begin{document}
\draft
 
\title{Correlation duality relations for the
($N_{\alpha}, N_{\beta}$) model}


 \author{F. Y. Wu and Wentao T. Lu}
\address{Department of Physics,
	 Northeastern University, Boston, Massachusetts 02115}

\maketitle

\begin{abstract}
Duality relations for the correlation functions of $n$ sites
on the boundary of   a planar lattice are derived for the
$(N_\a, N_\b)$ model of Domany and Riedel for $n=2,3$.
Our result holds for arbitrary lattices
which can have nonuniform interactions.
 
\end{abstract}

\vskip 1cm

\section{Introduction}
The duality relation for the Potts model 
 \cite{potts,wuPotts}
is an identity
 \cite{wuwang}  relating the partition functions of a Potts model
on a planar lattice with that of its dual.
Very recently,
 one of us \cite{wu} has  extended the  duality
consideration to Potts correlation functions.  Specifically, it was established that
 certain  duality relations exist 
for correlation functions of any $n$ 
Potts spins sitting on the boundary of a planar graph.
   Explicit expressions for the  duality relation were then obtained for
$n=2,3$.
While for $n=2$ the relation generalizes a known
expression \cite{watson,zia} for the 2-point correlation function of
the Ising model,
the general $n$ results are new.
 It was also stated   in \cite{wu} that the formulation can
be extended in a straightforward fashion to $n\geq 4$,  and to the
more general ($N_\a, N_\b$) model of Domany and Riedel \cite{domany}.
 
Jacobsen \cite{jacobsen} has since pointed out that the  procedure
as described in \cite{wu} does not extend straightforwardly
 to   $n\geq 4$.  But Wu and Huang \cite{wuhuang} subsequently showed 
that the procedure can be made to work, provided
that one invokes certain  correlation identities
which have not previously been known.
   In  this paper we address the other extension of the
formulation mention in \cite{wu}, namely, the extension to the
($N_\a, N_\b$) model.  Here, we report results for $n=2,3$.
 
\section{The ($N_\a, N_\b$) model}
 The ($N_\a, N_\b$) model \cite{domany} is an $N_\a N_\b$-state
spin model, where $N_\a$ and $ N_\b$ are positive integers.
For brevity we shall rename
\begin{equation}
q\equiv N_\a , \hskip 1cm \quad q' \equiv N_\b.
 \end{equation}
The model
is most conveniently visualized as being represented by
placing two Potts spins at each site.
At each site of  a lattice, or more generally a graph, $G$ of $N$ sites,
  one places two Potts spins  of $q$ and $q'$
states,  respectively.  Denote the spins states
by $\s=0,1,\cdots,q-1$ and $\t=0,1,\cdots,q'-1$.
 Then, the interaction energy $E(\s, \t; \s', \t' )$
between two sites in spin states
$\s, \t$ and $\s', \t'$ is
 given by
\bea
-E(\s, \t; \s', \t' )/kT& =& K_{00} \d_{\s, \s'} \d_{\t,\t'} + K_{01}
 \d_{\s,\s'}(1- \d_{\t, \t'}) \nonumber \\
&&\hskip .5cm +\ K_{10} (1-\d_{\s, \s'})
 \d_{\t,\t'} +K_{11}
    (1-\d_{\s, \s'})(1- \d_{\t,\t'}), \label{inter}
\eea
where $k$ is the Boltzmann constant, $T$ the temperature,
and $\d$ the Kronecker delta function.
The $(N_\a, N_\b)$ model becomes the Ashkin-Teller model \cite{at}
when $N_\a = N_\b =2$.

The partition function of the  $(N_\a, N_\b)$ model is  given by
\be
Z= \sum_{\s_i} \sum _{\t_i}\prod_{\rm edges} {\rm exp} \bigl[
       -E(\s_i, \t_i; \s_j, \t_j )/kT\bigr],  \label{part}
 \ee
where the product is taken over all pairs connected by edges
in $G$.
 
Consider $n$ sites on the boundary of ${\cal L}$.
The probability that the $n$ sites will be in specific definite states
$\{ \s_1, \t_1\}, \{ \s_2, \t_2\},\cdots
\{ \s_n, \t_n\}$ is given by
\be
  P(\s_1, \s_2, \cdots, \s_n \vert
\t_1, \t_2, \cdots, \t_n   ) = Z_{\s_1, \s_2, \cdots, \s_n ;
\t_1, \t_2, \cdots, \t_n} / Z, \label{prob}
\ee
where 
$Z_{\s_1, \s_2, \cdots, \s_n ;
\t_1, \t_2, \cdots, \t_n}$ is the partial partition function defined by
(\ref{part}) with the $n$ sites in definite states. 
   Then, following \cite{wu}, one can
define an $n$-point correlation
\be
\Gamma_n  \equiv (qq')^n P(\s, \s, \cdots, \s \vert
\t, \t, \cdots, \t  )  -1 ,  \label{ncorrelation}
\ee
where $ P(\s, \s, \cdots, \s \vert
\t, \t, \cdots, \t  ) $ is the probability that all $n$ sites are in
the same state.  Clearly,
$\Gamma_n$  vanishes identically when there is no correlation
between the $n$ sites. 

\section{Duality relation}
The partition function (\ref{part}) possess a duality relation \cite{wuwang,domany}.
Let $Z^*$ be the partition function of
an ($N_\a,N_\b$) model on the dual of $G$
which has $N^*$ sites with interactions also given by (\ref{inter}),
but with $K_{mn}$ replaced by $K^*_{mn},\ m,n=0,1$.
Further introduce Boltzmann factors $u_{mn} = e^{K_{mn}} $
and $u_{mn}^* = e^{K_{mn}^*}$.  Then, 
it is well-known \cite{wuwang,domany} that 
the duality relation assumes the form
\be
Z(u_{00},u_{01},u_{10},u_{11}) = ( qq'C)
Z^*(u_{00}^*,u_{01}^*,u_{10}^*,u_{11}^*), \label{dual}
\ee
where $C= (qq')^{-N^*}$ and
\bea
u_{00}^* &=&u_{00} +(q-1)u_{10}  +(q'-1)u_{01}+(q-1)(q'-1)u_{11} \nonumber \\
u_{01}^* &=&u_{00} -u_{01}  +(q-1)(u_{10}-u_{11}) \nonumber \\
u_{10}^* &=&u_{00} -u_{10}  +(q'-1)(u_{01}-u_{11}) \nonumber \\
u_{11}^* &=&u_{00} -u_{01}  -u_{10}+u_{11}. \label{dualinter}
\eea
Note that the transformation (\ref{dualinter}) can be written more compactly as
\be
{\bf u}^* ={\bf T}_2(q)\cdot {\bf u} \cdot {\tilde {\bf T}_2}(q') ,
\ee
where 
\be
{ {\bf u}} =\pmatrix{u_{00} & u_{01} \cr
            u_{10}&  u_{11}\cr } , \hskip .5cm
{\bf u}^* =\pmatrix{u^*_{00} & u^*_{01} \cr
            u^*_{10}&  u^*_{11}\cr } , \hskip .5cm
{\bf T}_2(q) =\pmatrix {1 & q-1 \cr 1 & -1 \cr}. \label{t2}
\ee
and ${\tilde {\bf T}}$ is the transpose of ${\bf T}$.
 We shall refer to (\ref{dual})  as the fundamental
duality relation which applies to arbitrary graph $G$ with
arbitrary (nonuniform) edge interactions.  As discussed in \cite{wu},
the duality relations for correlation functions  are
most conveniently obtained by 
applications of this fundamental duality relation.

\section{The 2-point correlation function} 
In this section  we 
 consider the duality for the 2-point correlation functions between
any two spins at sites $i$ and $j$ on the boundary of $G$.
Following \cite{wu,wuhuang}, this is done by applying
(\ref{dual}) to an auxiliary graph
(lattice). The auxiliary graph is obtained from $G$ by connecting
sites $i$ and $j$ with an auxiliary edge as shown in Fig. 1(a). Let
$u_{mn}$ and $u^*_{mn}$ be the respective
Boltzmann factors associated with this edge and 
its dual 
 related by (\ref{dualinter}).
 Applying the fundamental duality to the auxiliary 
graph of Fig. 1(a), we arrive at 
(\ref{dual}) in the form
\be
Z_{\rm aux} =  C Z^*_{\rm aux},
\ee
where we have used the fact that the dual of the auxiliary graph has $N^*+1$ sites, and
\bea
Z_{\rm aux} &=&qq'[u_{00}Z_{00;00}+(q-1)u_{10}Z_{01;00}
	+(q'-1)u_{01}Z_{00;01}+(q-1)(q'-1)u_{11}Z_{01;01}] \nonumber \\
Z^*_{\rm aux} &=&qq'[u^*_{00}Z^*_{00;00}+(q-1)u^*_{10}Z^*_{01;00}
	+(q'-1)u^*_{01}Z^*_{00;01}+(q-1)(q'-1)u^*_{11}Z^*_{01;01}]. \label{auxil}
\eea
Note that in writing down (\ref{auxil}) 
we have made use of the degeneracy $Z_{00;00}=Z_{\a\a;\b\b}$;
$Z_{00;01} = Z_{\a\a;\a\b}, \a\not= \b$; etc.

Substituting (\ref{dualinter}) into (\ref{auxil}), the dual relation
(\ref{dual}) becomes   linear in
both $u_{mn}$ and $Z_{\s_i\s_j,\t_i\t_j}$. 
It is now a simple matter to equate the coefficients of 
the 4 $u_{mn}$, and obtain 
 \bea
Z_{00;00}&=&C[Z^*_{00;00}+(q-1)Z^*_{01;00}+(q'-1)(Z^*_{00;01}+(q-1)Z^*_{01;01})] \nonumber \\
Z_{01;00}&=&C[Z^*_{00;00}-Z^*_{01;00}+(q'-1)(Z^*_{00;01}-Z^*_{01;01})]\nonumber \\
Z_{00;01}&=&C[Z^*_{00;00}-Z^*_{00;01}+(q-1)(Z^*_{01;00} -Z^*_{01;01})] \nonumber \\
Z_{01;01}&=&C[Z^*_{00;00}-Z^*_{01;00}-Z^*_{00;01}+Z^*_{01;01}]. \label{a}
\eea
This relation 
can be written more compactly as
\be
{\bf Z}_2 =C{\bf T}_2(q)\cdot {\bf Z}_2^* \cdot 
{\tilde {\bf T}_2}(q'), \label{z}
\ee
where 
${\bf Z}_2$ is a $2\times 2$ matrix 
with $(\b\b')$-th element $Z_{0\b;0\b'}$,
and similarly for ${\bf Z}_2^*$.
This is the desired duality relation for the 2-point correlation function.

To compute the  correlation (\ref{ncorrelation}) for $n=2$, we  use 
(\ref{prob}) and note that one can apply
the fundamental duality relation (\ref{dual}) to ${\cal L}$
to write  the partition function $Z$ 
in the form of 
\be 
Z = qq'C Z^* = (qq')^2 C Z^*_{00;00}. \label{zz}
\ee
Substituting (\ref{a}) and (\ref{zz}) into (\ref{ncorrelation}),
one is led to the result
\bea
\Gamma_2 &=& (q-1) p_{01;00} + (q'-1) p_{00;01} +(q-1)(q'-1) p_{01;01}  \nonumber \\
&=& {\bf v}_2(q)\cdot{\bf p}_2\cdot{\tilde {\bf v}_2}(q')-1, \label{G2}
\eea
where 
\be 
p_{\a\b;\a'\b'} \equiv Z^*_{\a\b;\a'\b'}/ Z^*_{00;00} , \hskip 1cm \a,\b,\a',\b' = 0,1, \label{gamma2}
\ee
with
\bea
{\bf v}_2(q)&=&\pmatrix{ 1, & q-1\cr} , \nonumber \\
{\bf p}_2&=&\biggl({1\over {Z^*_{00;00}}}\biggr){\bf Z}^*_2.
\eea
The expression (\ref{gamma2}) generalizes the $q'=1$ result for
the Potts model \cite{wu}.
    
\section{THE 3-POINT CORRELATION FUNCTION}
For the 3-point correlation functions, we apply the fundamental duality relation
to the auxiliary graph shown in Fig. 1(b), and extract from the resulting expression
the desired duality relation.
Now for $n=3$
there are 25 independent 3-point correlation functions
and the algebra tends to be involved.
The procedure of \cite{wu} can nevertheless 
be carried through, and the result
can be expressed compactly as suggested by (\ref{zz}) as follows.
 
Introduce a $5\times 5 $ matrix  ${\bf Z}_3$ whose elements are the 
3-point correlation functions
\be
{\bf Z}_3=\pmatrix{Z_{000;000}& Z_{000;001}& Z_{000;010}& Z_{000;100}&Z_{000;012}\cr
                 Z_{001;000}& Z_{001;001}& Z_{001;010}& Z_{001;100}&Z_{001;012}\cr
                 Z_{010;000}& Z_{010;001}& Z_{010;010}& Z_{010;100}&Z_{010;012}\cr
                 Z_{100;000}& Z_{100;001}& Z_{100;010}& Z_{100;100}&Z_{100;012}\cr
                 Z_{012;000}& Z_{012;001}& Z_{012;010}& Z_{012;100}&Z_{012;012}\cr},
\ee
and similarly for ${\bf Z}_3^*$.
Then, we find 
\be
{\bf Z}_3= \biggl({C\over {qq'}} \biggr)
{\bf T}_3(q)\cdot {\bf Z}^*_3\cdot {\tilde {\bf T}_3}(q') \label{z3}
\ee
where 
\be
{\bf T}_3(q)=
\pmatrix{ 1&q-1 & q-1 & q-1& (q-1)(q-2)\cr
          1&  -1& q-1&  -1&  -(q-2)\cr
          1&  -1&  -1& q-1&  -(q-2)\cr
          1& q-1&  -1&  -1&  -(q-2)\cr
          1&  -1&  -1&  -1&     2\cr}.
\ee
 
The 3-point correlation (\ref{ncorrelation}) can be computed straightforwardly
using (\ref{z3}) and the identity
\be
Z= (qq')^2 C Z^*_{000;000}.
\ee
This leads to the compact expression which extends (\ref{G2}),
\be
\Gamma_3={\bf v}_3(q)\cdot{\bf p}_3\cdot{\tilde {\bf v}_3}(q')-1,
\ee
where the row vector ${\bf v}_3$ and the $5\times 5$ matrix {\bf p}$_3$ are given by
\bea
{\bf v}_3(q)&=&\pmatrix{1,& q-1,& q-1,& q-1,&(q-1)(q-2)\cr}, \nonumber \\
{\bf p}_3&=&\biggl({1\over {Z^*_{000;000}}}\biggr){\bf Z}^*_3.
\eea
\section{Summary and Acknowledgement}
We have obtained the duality relations for the 2- and 3-site boundary correlation
functions for the $(N_\a, N_\b)$ model, which includes the 
$N_\a =N_\b =2$ Ashkin-Teller model
as a special case.  Explicit expressions
are also obtained for the  2- and 3-point correlations.  Our results are
presented in  compact forms which are suggestive of possible  
extensions to higher correlations
and to the chiral Potts model \cite{chiral}.  These extensions
  will be reported elsewhere \cite{luwu}.

This work has been supported in part by National Science Foundation
Grant No. DMR-9614170.

\begin{center}

{\bf Figure captions}

\end{center}

\noindent
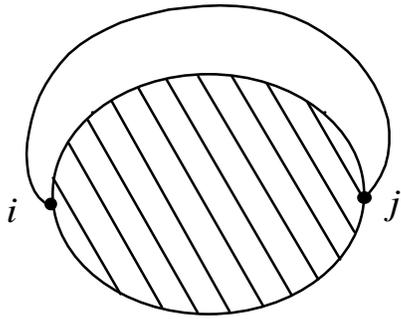
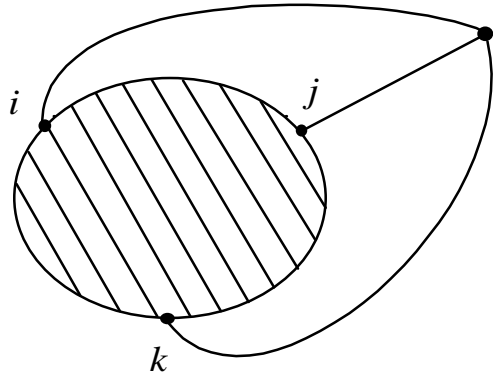
Fig. 1.
(a) The auxiliary graph for $n=2$.
(b) The auxiliary graph for $n=3$.

\end{document}